\documentclass[a4paper,11pt]{article}
\pdfoutput=1
\usepackage{jheppub}

\title{Universal chiral conductivities for low temperature holographic superfluids}

\author[a,b]{I. Amado,}
\author[a]{N. Lisker}
\author[a]{and A. Yarom}

\affiliation[a]{Department of Physics, Technion, Haifa 32000, Israel}
\affiliation[b]{Department of Mathematics and Physics, University of Haifa, Qiryat Tivon 36006, Israel}

\emailAdd{iamado}
\emailAdd{nlisker}
\emailAdd{ayarom@physics.technion.ac.il}

\abstract{We argue that the chiral conductivities of generic $s$-wave holographic superfluids, whose broken $U(1)$ symmetry is anomalous, exhibit universal behavior at low temperatures. The universal behavior we argue for is independent of the details of the bulk action and on the emergent geometry deep in the bulk interior at low temperatures. Our results are contrasted against general expectations based on an analysis of the entropy current in superfluids.}

\begin{document}

\maketitle

\section{Introduction and summary}

Recent developments have thrust forward our understanding of the role of anomalies in hydrodynamic and thermal states. Somewhat surprisingly, anomalies enter the hydrodynamic constitutive relations in unexpected ways, allowing for transport phenomenon associated with the response of a fluid to a magnetic field and vorticity \cite{Vilenkin:1980fu,Erdmenger:2008rm,Banerjee:2008th,Son:2009tf,Neiman:2010zi,Landsteiner:2011cp}. A fluid possessing a $U(1)^3$ anomaly or a mixed anomaly will linearly respond to even small vorticities whereas a fluid for which anomalies are absent will not. A full study of the role of anomalies in normal fluids has recently been discussed in \cite{Jensen:2013kka,Jensen:2013rga}.

In what follows we present a study of the role of anomalies in superfluids. Superfluids may be thought of as fluids for which a global $U(1)$ symmetry is spontaneously broken. This phenomenon manifests itself in terms of an extra superfluid degree of freedom, the Goldstone boson. Like normal fluids, superfluids may also respond to vorticity (of the normal fluid component) and a magnetic field. However, the standard analysis of superfluids poses no restrictions on the response parameters and the response may, or may not, depend on the anomalies present in the theory. This unmistakable difference between normal fluids, for which the response to vorticity and magnetic field is fully controlled by the anomalies, and superfluids, for which the response to vorticity and magnetic field is not controlled by the anomalies, is not surprising. The extra goldstone degree of freedom available to superfluids changes its dynamics to the hilt \cite{LL6,Putterman}. Indeed,  canonical methods used to study superfluids give very little data as to the value of the response of the superfluid to vorticity or a magnetic field \cite{Lublinsky:2009wr,Bhattacharya:2011eea,Lin:2011aa,Bhattacharya:2011tra,Neiman:2011mj,Bhattacharyya:2012xi,Chapman:2013qpa}. Additionally, holographic methods, as used so far \cite{Bhattacharya:2011tra}, have also not provided clear predictions for the aforementioned response parameters.

Here we extend the analysis of \cite{Bhattacharya:2011tra} where, among other things, parity violating transport of holographic superfluids was studied. In a holographic framework the response of the superfluid to vorticity or a magnetic field is, in general, model dependent. However, as we will shortly argue, if parity is broken due to an anomaly then at low temperatures the parity breaking response parameters of the (s-wave) superfluid are completely fixed by the anomalies of the theory and do not depend on the particular details of the bulk Lagrangian. This feature is somewhat unexpected especially since it is oblivious to the bulk geometry emerging in the infrared at low temperatures. That is, the universal value we obtain for the response to vorticity and magnetic field, is the same whether the geometry that emerges at zero temperature has conformal symmetry or Lifshitz symmetry. Some other setups where transport properties of holographic theories are universal can be found in \cite{Policastro:2001yc,Buchel:2003tz,Kovtun:2004de,Buchel:2004qq,Haack:2008xx,Shaverin:2012kv}.

To make our claim more precise, let us introduce some notation. A normal fluid in $3+1$ dimensions with a single conserved $U(1)$ charge may be characterized by a velocity field $u^{\mu}$, a temperature field $T$ and a chemical potential $\mu$. The existence of a hydrodynamic state implies that the stress tensor and conserved charge current may be expressed as a local function of the hydrodynamic fields $u^{\mu}$, $T$ and $\mu$ and their gradients. These expressions are referred to as constitutive relations. Assuming that variations of the hydrodynamic fields are small compared to the mean free path, the constitutive relations may be expressed as a series expansion in gradients of $u^{\mu}$, $T$ and $\mu$ (See e.g., \cite{Baier:2007ix,Bhattacharyya:2008jc} for an extensive discussion). For a normal fluid, with a $U(1)^3$ anomaly and with a particular choice of out of equilibrium definition of temperature, velocity and chemical potential, (i.e., the Landau frame \cite{LL6}) one finds by a standard analysis \cite{Son:2009tf,Neiman:2010zi} that
\begin{equation}
\label{E:Jnormal}
	J^{\mu} = \rho u^{\mu} + \kappa \left(T^{-1} F^{\mu\nu}u_{\nu} - P^{\mu\nu}\partial_{\nu} \frac{\mu}{T}\right)+\tilde\kappa_{\omega}\omega^{\mu} + \tilde{\kappa}_{B}B^{\mu}\,,
\end{equation}
where $\rho(T,\mu)$ is the charge density, $\kappa(T,\mu)$ is the thermal conductivity, $F^{\mu\nu}$ is the external field strength, $P^{\mu\nu} = \eta^{\mu\nu}+u^{\mu}u^{\nu}$ is the projection matrix orthogonal to the velocity field, $\omega^{\mu} = \epsilon^{\mu\nu\rho\sigma}u_{\nu}\partial_{\rho} u_{\sigma}$ is the vorticity and $B^{\mu} = \frac{1}{2}\epsilon^{\mu\nu\rho\sigma}u_{\nu}F_{\rho\sigma}$ is the magnetic field as seen in the rest frame of a fluid element. In what follows we will refer to the last two terms on the right hand side of \eqref{E:Jnormal} as the parity violating terms and the remaining terms on the right hand side of \eqref{E:Jnormal} as the parity preserving terms. While the conductivity $\kappa$ can take any positive value, the conductivities $\tilde{\kappa}$ must satisfy
\begin{equation}
\label{E:kappatildenormal}
	\tilde{\kappa}_{\omega} = c \left(\mu^2 - \frac{2}{3} \frac{\rho}{\epsilon+P} \mu^3 \right)\,,
	\qquad
	\tilde{\kappa}_{B} = c\left(\mu - \frac{1}{2} \frac{\rho}{\epsilon+P} \mu^2 \right)\,,
\end{equation}
with $\epsilon$ and $P$ the energy density and pressure respectively. Here $c$ specifies the strength of the anomaly, i.e., the (non-)conservation law for the current $J^{\mu}$ is given by \cite{Sahoo:2010sp}:\footnote{In this work we consider only a $U(1)^3$ anomaly. Mixed anomalies also play an important role in the hydrodynamic response of $3+1$ dimensional theories \cite{Landsteiner:2011cp,Golkar:2012kb,Jensen:2012kj,Jensen:2013kka,Jensen:2013rga} but will not be discussed in the current work.}
\begin{equation}
	\partial_{\mu}J^{\mu} = -\frac{c}{8} \epsilon^{\mu\nu\rho\sigma}F_{\mu\nu}F_{\rho\sigma}\,.
\end{equation}
In what follows we will (with a slight abuse of language) refer to the coefficients $\tilde\kappa$ as chiral conductivities.

While the anomaly contributes to parity breaking terms in the current, as given by \eqref{E:Jnormal}, with our current choice of out of equilibrium definitions for the velocity field, temperature and chemical potential it does not modify the constitutive relations for the stress tensor. It does, however, modify the entropy current. An entropy current is a current which in thermodynamic equilibrium is given by the entropy density boosted to the fluid center of mass velocity and whose divergence is non negative for any solution to the equations of motion \cite{LL6,Bhattacharya:2011tra}. In a hydrodynamic setting, and with our current conventions, it is given, to first order in derivatives, by
\begin{equation}
\label{E:Jsnormal}
	J_s^{\mu} = s u^{\mu} - \frac{\mu}{T}\left(J^{\mu} - \rho u^{\mu}\right) + \sigma_{\omega} \omega^{\mu} + \sigma_B B^{\mu}\,.
\end{equation}
The coefficients $\sigma_{\omega}$ and $\sigma_B$ are referred to as the non canonical contributions to the entropy current and are given by
\begin{equation}
\label{E:sigmatildenormal}	\sigma_{\omega} = c \frac{\mu^3}{3 T}\,, \qquad 
	\sigma_{B} = c\frac{\mu^2}{2 T}\,.
\end{equation}
See \cite{Son:2009tf} for details.

For a superfluid with small superfluid velocity, the same techniques that lead to \eqref{E:Jnormal} give us 
\begin{equation}
	J^{\mu} =  \left(\substack{ \hbox{parity preserving} \\ \hbox{terms}} \right) +\tilde\kappa_{\omega}\omega^{\mu} + \tilde{\kappa}_{B}B^{\mu}\,.
\end{equation}
However, following the analysis in  \cite{Bhattacharya:2011tra,Bhattacharyya:2012xi,Chapman:2013qpa}, as opposed to relations of the form \eqref{E:kappatildenormal}, the chiral conductivities are not fixed by the anomaly. Whether the underlying theory is anomalous or not, as long as parity is broken we expect $\tilde{\kappa}_{\omega}$ and  $\tilde{\kappa}_B$ to be non zero. Similarly, the parity odd contributions to the entropy current are given by
\begin{equation}
	J_s^{\mu} = \left(\substack{ \hbox{parity preserving} \\ \hbox{terms}} \right) + \left( {\sigma}_{\omega} - \frac{\mu}{T} \tilde{\kappa}_{\omega} \right)\omega^{\mu} +\left( {\sigma}_{B} - \frac{\mu}{T} \tilde{\kappa}_B\right) B^{\mu}\,,
\end{equation}
where the coefficients $\sigma_B$ and $\sigma_{\omega}$ must satisfy 
\begin{equation}
	\frac{1}{2}\sigma_{\omega} - \mu \sigma_B = -\frac{c\mu^3}{3 T}\,,
\end{equation}
but are otherwise unconstrained \cite{Bhattacharya:2011tra}. Nonetheless, we claim in what follows that in generic holographic superfluids the chiral terms will asymptote to the values
\begin{equation}
\label{E:asymptotics}
	\tilde{\kappa}_{\omega} = 0\,,
	\qquad
	\tilde{\kappa}_B = \frac{c}{3}\mu\,,
	\qquad
	{\sigma}_{\omega}  = 0\,,
	\qquad
	{\sigma}_B = \frac{\mu}{T} \tilde{\kappa}_B = \frac{c}{3} \frac{\mu^2}{T}
\end{equation}
at low temperatures.\footnote{The values quoted in \eqref{E:asymptotics} are the ones associated with the covariant current. The chiral conductivities for the consistent current vanish. For a detailed discussion on consistent and covariant currents see \cite{Bardeen:1984pm,Jensen:2013kka}. We thank K. Jensen for discussions on this point.} In other words, chiral transport coefficients in holographic superfluids with anomalies approach a universal value at low temperatures.  In order to estimate the rate at which the chiral conductivities converge to their low temperature values \eqref{E:asymptotics} we resort to numerics. We find that superfluids whose geometry approaches an AdS to AdS domain wall solution at low  temperatures have good convergence properties but that geometries whose low temperature behavior is asymptotically Lifshitz in the deep interior must be very cold in order for \eqref{E:asymptotics} to hold to good accuracy. 

Our paper is organized as follows, in section \ref{S:Review} we review the results of \cite{Bhattacharya:2011tra} where integral expressions for $\tilde{\kappa}_B$, $\tilde{\kappa}_{\omega}$, $\sigma_{\omega}$ and $\sigma_{B}$ have been obtained for generic $s$-wave superfluids. Using these integral expressions (see equation \eqref{E:oddRules}) we argue in section \ref{S:lowT} that at low enough temperatures $\tilde{\kappa}_B$, $\tilde{\kappa}_{\omega}$, $\sigma_{\omega}$ and $\sigma_{B}$ should asymptote to \eqref{E:asymptotics} regardless of the emergent geometry in the deep interior. We complement our analytic findings with numerical solutions to the equations of motion in section \ref{S:Numerics} where we also discuss the rate at which the chiral conductivities approach their universal values \eqref{E:asymptotics}. 

\section{Review of the results of \cite{Bhattacharya:2011tra}}
\label{S:Review}


The minimal holographic realization of a superfluid consists of an Abelian gauge field $A_M$ and a charged scalar field $\psi$ living on an asymptotically AdS geometry \cite{Gubser:2008px,Hartnoll:2008kx,Hartnoll:2008vx}. Spontaneous breaking of a global $U(1)$ symmetry in the boundary theory is triggered by the condensation of the charged scalar field in the bulk. The most general five dimensional two-derivative bulk action which includes gravity, a $U(1)$ gauge field and a charged scalar field is given by
\begin{equation}
\label{E:action}
	S= S_{\rm EH} + S_{matter} + S_{\rm CS}\,,
\end{equation}
where 
\begin{align}
\begin{split}
	S_{\rm EH} &= \frac{1}{2\kappa^2} \int d^5x\sqrt{-g} \left(R+12\right)\,,\nonumber\\
	S_{matter} &= \frac{1}{2\kappa^2} \int d^5x\sqrt{-g} \left(-\frac{1}{4} V_F(|\psi|) F_{MN} F^{MN} - V_\psi (|\psi|)(D_M \psi) (D^M \psi)^* - V(|\psi|)\right)\,,\nonumber\\
	S_{\rm CS} &= \frac{c}{24} \int d^5x\sqrt{-g} \epsilon^{MNPQR} A_M F_{NP} F_{QR}\,,
\end{split}
\end{align}
with $\epsilon^{01234}=1/\sqrt{-g}$, $D_M=\partial_M - i q A_M$, $V_F (0) = V_\psi (0) =1$ and $V(0) =0$. Capital latin indices run from $0$ to $4$. The parity odd topological term $S_{\rm CS}$ implies that the dual field theory posseses a $U(1)^3$ anomaly. The parameter $c$ which characterizes the Chern-Simons term is identified with the strength of the anomaly. 
%
The equations of motion which follow from the action \eqref{E:action} are
\begin{align}
\begin{split}
\label{E:EOMs}
	0 &= \frac{1}{\sqrt{-g}} D_M \left(\sqrt{-g} V_\psi D^{M}\psi\right)-\frac{\partial V(|\psi|)}{\partial \psi^*}-\frac{1}{4} \frac{\partial V_F(|\psi|)}{\partial \psi^*}  F^2-\frac{\partial V_\psi(|\psi|)}{\partial \psi^*} |D\psi|^2\,,\\
	0 &= \frac{1}{\sqrt{-g}} \partial_N \left(\sqrt{-g}V_F F^{NM}\right) + \frac{\kappa^2 c}{4} \epsilon^{MABCD} F_{AB} F_{CD} + i q V_\psi\left( \psi (D^M \psi)^* - \psi^* (D^M \psi)\right) \,,\\
	0&= R_{MN} -\frac{1}{2} (R+12) g_{MN} - \mathcal{T}_{MN}\,,
\end{split}
\end{align}
with
\begin{align}
\begin{split}
 \mathcal{T}_{MN} =& -\frac{1}{2} g_{MN} \left(\frac{1}{4} V_F F_{AB} F^{AB} + V_\psi  |D \psi  |^2 +V\right) +\frac{1}{2} V_F F_{MA} F_N\,^{A} \nonumber\\
 &+\frac{1}{2} V_\psi\left((D_M \psi)(D_N \psi)^*+(D_N \psi)(D_M \psi)^*\right) \,.
\end{split}
\end{align}
Various numerical solutions to the equations of motion \eqref{E:EOMs} have been studied in the literature (see, e.g., \cite{Herzog:2009md,Gubser:2009qm,Arean:2010wu}). 

In \cite{Bhattacharya:2011tra} a closed form expression for the parity odd transport coefficients $\tilde{\kappa}_{\omega}$, $\tilde{\kappa}_B$, $\sigma_{\omega}$ and $\sigma_B$ has been obtained. In what follows we sketch-out the computation of \cite{Bhattacharya:2011tra} referring the interested reader to the latter for a full derivation. We start with an ansatz
\begin{align}
\begin{split}
\label{E:ansatz}
	ds^2 =& -r^2 f(r) dt^2 + r^2 d\vec{x}^2 + 2 h(r) dt dr \,,\\
	\psi =& \varrho (r) e^{i q \varphi(r)} \,,\\
	A_M =& \left(A_0 (r),0,0,0,A_4 (r) \right)
\end{split}
\end{align}
describing a stationary superfluid in the dual field theory, and replace the gauge field with the gauge invariant combination
\begin{equation}
G_M = A_M -\partial_M \varphi\,.
\end{equation}

Solutions to the equations of motion \eqref{E:EOMs} which describe a superfluid at non zero temperature are characterized by an event horizon which manifests itself as a simple zero of $f$ and $G_0$ at a fixed value of the radial coordinate $r_h$. The Hawking temperature and Bekenstein entropy density of the resulting black hole are given by
\begin{equation}
\label{E:BH}
T=\frac{r_h^2 f'(r_h)}{4 \pi h(r_h)}\,,\qquad s=\frac{2 \pi r_h^3}{\kappa^2}\,.
\end{equation}
The latter are also the temperature and entropy density of the dual field theory. Other thermodynamic properties of the dual field theory may be read off of the near boundary (large $r$) expansion of the bulk fields, viz., 
\begin{align}
\begin{split}
	\label{E:UV}
	f&= 1 - \frac{2 \kappa^2 P}{r^4} +\mathcal{O}(r^{-5})\,,\qquad
	h= 1 - \frac{\Delta C_\Delta^2 |\left\langle O_\psi\right\rangle |^2}{6 r^{2 \Delta}}+\mathcal{O}(r^{-2\Delta -2})\,,\\
	\varrho &=\frac{C_\Delta |\left\langle O_\psi\right\rangle |}{r^\Delta} + \mathcal{O}(r^{\Delta -2})\,,\qquad
	G_0 = \mu - \frac{\kappa^2 \rho_t}{r^2} + \mathcal{O}(r^{-3})\,,
\end{split}
\end{align}
where $\Delta$ is the conformal dimension of the operator $O_\psi$ dual to the scalar $\psi$ and $C_\Delta$ is a real constant which can be determined but is irrelevant for our current discussion. The parameters $\mu$ and $\rho_t$ correspond to the chemical potential and the total charge density on the superfluid phase, respectively. 

The gravitational manifestation of the Gibbs Duhem relation between entropy, temperature, energy density and pressure follows from the existence of a Noether charge \cite{Gubser:2009qf,Gubser:2009cg}
\begin{equation}
\label{E:Q1}
	2 \kappa^2 Q_1 = \frac{r^5 f'-r^3 V_F G_0 G'_0}{h}\,.
\end{equation}
It is straightforward to check that $\partial_r Q_1=0$ under the equations of motion and that using \eqref{E:BH} and \eqref{E:UV}
\begin{equation}
	Q_1 = s T = 4 P - \mu \rho_t\,.
\end{equation}

In order to obtain a non stationary superfluid with small superfluid velocity one needs to introduce linear perturbations of the gauge field $G_i$ and metric components $g_{ti}$ where $i=1,\ldots,3$ \cite{Herzog:2008he,Herzog:2009md}. Denoting the Goldstone boson in the boundary theory by $\phi$, we may write
\begin{equation}
G_i = - g(r) \partial_i \phi\,,\qquad g_{ti} = - r^2 \gamma (r) \partial_i \phi\,.
\end{equation}
The linearized equations of motion for the fluctuations $\gamma$ and $g$ can be written as total derivatives. After integrating these equations once we find
\begin{subequations}
\label{E:ggamma}
\begin{align}
	\label{E:Q2}2 \kappa^2 Q_2 &= \frac{r^5 \gamma' + r^3 V_F g G'_0}{h} \,,\\
	\label{E:eqg}Q_3 + 2 \kappa^2 f Q_2 &= 2 \kappa^2 \gamma Q_1 + \frac{f r^3 V_F \left(g G'_0 - g' G_0\right)}{h}\,,
\end{align}
\end{subequations}
where $Q_2$ and $Q_3$ are integration constants. The presence of an event horizon at $r=r_h$ implies that $\gamma(r_h)=0$ and hence that $Q_3=0$. 

When the temperature of the superfluid is non vanishing then only a fraction of the charge condenses and the total charge density of the system may be thought of as a sum of a superfluid (condensate) charge density and a normal charge density.
The charge density of the normal (uncondensed) component, $\rho$, can be read off of the near boundary (large $r$) expansion of $g$ and $\gamma$,
\begin{align}
\begin{split}
\gamma &= \frac{1}{2} \frac{(\rho_t - \rho)\kappa^2}{r^4} + \mathcal{O} (r^{-5})\,,\\
\label{E:asymg}g &= 1- \frac{(\rho_t - \rho) \kappa^2}{\mu r^2} + \mathcal{O} (r^{-3})\,.
\end{split}
\end{align}
Inserting \eqref{E:asymg} into \eqref{E:Q2} we find that
\begin{equation}
Q_2 = \rho \,.
\end{equation}
Carrying out a standard bulk to boundary mapping (see e.g., \cite{Herzog:2008he,Herzog:2011ec,Bhattacharya:2011eea}) one finds that
\begin{equation}
	J_{\mu} = \rho u_{\mu} +\frac{\rho_t-\rho}{\mu} \partial_{\mu}\phi
\end{equation}
as expected for a superfluid at leading order in gradients moving at a small superfluid velocity \cite{LL6,Herzog:2008he,Herzog:2009md,Herzog:2011ec,Bhattacharya:2011tra}.

In \cite{Bhattacharya:2011tra} a straightforward though somewhat tedious computation shows that the parity odd transport coefficients associated with this theory are given by definite integrals over the bulk fields $g$ and $G_0$. The computation involves carrying out the fluid-gravity algorithm for superfluids \cite{Herzog:2011ec,Bhattacharya:2011eea}, or, in other words, perturbing the superconducting black hole solution described above and studying the tensor, vector and scalar modes associated with the slowly space-time varying perturbations . The coefficients $\tilde{\kappa}_{\omega}$ $\tilde{\kappa}_B$, $\sigma_{\omega}$ and $\sigma_B$ are associated with parity odd vector modes whose equations of motion decouple from the scalar and tensor modes and can be solved for implicitly in integral form. In particular, one finds that
\begin{subequations}
\label{E:oddRules}
\begin{align}
	\tilde\kappa_{B} &= c \int_{r_h}^{\infty}g^2 G_0' +R (G_0-g\mu)g G_0' dr \,,\\
	\tilde\kappa_\omega &= -2c \int_{r_h}^{\infty} (G_0-\mu g) g G_0' + R (G_0-\mu g)^2 G_0' dr \,,\\
	\sigma_{B} & = \frac{c}{T} \int_{r_h}^{\infty} g G_0 G_0' dr \,,\\
	\sigma_{\omega} & = -\frac{2 c}{T} \int_{r_h}^{\infty} (G_0-\mu g) G_0 G_0' dr \,,
\end{align}
with
\begin{equation}
	R = \frac{\rho}{4P-\mu(\rho_t-\rho)}\,.
\end{equation}
\end{subequations}

\section{Low temperature behavior}
\label{S:lowT}
The parity odd transport coefficients $\tilde{\kappa}_{\omega}$, $\tilde{\kappa}_B$, ${\sigma}_\omega$ and ${\sigma}_B$ are given by integrals of $G_0$ and $g$. At low temperatures or small $Q_1$ an approximate solution to \eqref{E:eqg} is given by
\begin{equation}
\label{E:Q20}
	\frac{\mu g}{G_0}= 1 + \mu \int_r^{\infty} \frac{2 \kappa^2 Q_2 h}{V_F(\psi) G_0^2 r'^3 } dr' + \mathcal{O}(Q_1)\,,
\end{equation}
where we have imposed that $g = 1$ at the boundary. In order for $g/G_0$ to remain finite at the horizon we need to set $Q_2$ to $0$. Thus, as one might expect, we find that at low temperatures the charge density of the normal component vanishes. Inserting $g = G_0/\mu$ into \eqref{E:oddRules} one finds that the integrands in \eqref{E:oddRules} become total derivatives leading to \eqref{E:asymptotics}.

That zero temperature is equivalent to zero normal charge density is the key feature of holographic superfluids which enables us to compute the zero temperature limit of $\tilde{\kappa}_\omega$, $\tilde{\kappa}_B$, $\sigma_{\omega}$ and $\sigma_B$. 
To understand how the latter condition is realized we follow the analysis of \cite{Gubser:2009cg,Horowitz:2009ij}. Consider the zero temperature limit of the black hole dual to the condensed phase. By construction, the asymptotic behavior of the fields at large values of $r$ approach an AdS solution,
\begin{equation}
\label{E:AdSUV}
	ds^2 = r^2 \left(-dt^2 + d\vec{x}^2\right) + 2dt dr\,. 
\end{equation}	
Deep in the interior of AdS, when $r$ is small, the solution to the equations of motion will approach a different stationary configuration where the magnitude of the scalar $|\psi|$ is a constant. We will refer to the full solution interpolating between the large $r$ (ultraviolet) asymptotically AdS regime and the small $r$ (infrared) regime as a domain wall. Within our setup \eqref{E:action} and \eqref{E:ansatz} there are two classes of infrared behavior \cite{Gubser:2009cg} whose precise form depends on the value of the scalar field in the deep interior.\footnote{%
In \cite{Gubser:2009cg} it was shown that, in $AdS{}_4$, assuming an isotropic space-time the infrared behavior of the dual CFT could have Lifshytz or conformal symmety. In this work we restrict ourselves to superfluids which are isotropic. Non isotropic solutions such as those mentioned in \cite{Nakamura:2009tf} or hard wall models for confinement as discussed in \cite{Karch:2006pv} will not be considered in this work---their general effect on the superfluid dynamics has yet to be investigated.}
If the scalar $\psi$ is at a minimum of the potential $V(\psi)$ then the infrared geometry is an AdS solution,
\begin{equation}
\label{E:AdSIR}
	ds^2=\frac{r^2}{L_{IR}^2} \left(-f_0 dt^2 + d\vec{x}^2\right)+2 \frac{\sqrt{f_0}}{L_{IR}} dr dt
\end{equation}	
together with $G_0=0$ and $g=\gamma=0$,
where
\begin{equation}
	L_{IR} = \sqrt{\frac{-12}{-12+V(\psi_{IR})}}\,,
\end{equation}
and $\psi_{IR}$ is the value of the scalar at the minimum of $V(\psi)$. Alternately $\psi$ can take values $\psi_0$ for which the geometry takes the Lifshitz form
\begin{align}
\begin{split}
\label{E:LifshitzIR}
	ds^2 &=  -\frac{z p_0^2 V_F(\psi_0)}{2(z-1)} r^{2z} dt^2 + r^2 d\vec{x}^2 + \frac{\sqrt{3} z p_0 V_F(\psi_0)}{q \psi_0 \sqrt{(z-1)V_{\psi}(\psi_0)}} r^{z-1} dr dt\,, \\
	G_0&=p_0 r^z\,,
\end{split}
\end{align}
where $z$ is given implicitly by
\begin{equation}
	V'(\psi_0) = 2 \frac{(z-1)V_\psi (\psi_0)}{z\,V_F (\psi_0)} q^2 \psi_0 \left(2+ \frac{\psi_0}{V_\psi (\psi_0)} V_\psi'(\psi_0)+ \frac{z\,\psi_0}{3\,V_F (\psi_0)} V_F' (\psi_0)\right)\,,
\end{equation}
and reality of the solution implies that $z \geq 1$.
For the Lifshitz solution the vector fluctuations satisfy
\begin{equation}
\label{E:Lifshitzfluc}
	g=g_0 r^z \qquad \hbox{and} \qquad \gamma = -\frac{z g_0 p_0 V_F(\psi_0)}{2 (z-1)} r^{2z-2}\,.
\end{equation}
We will refer to the solution interpolating from \eqref{E:AdSUV} to \eqref{E:AdSIR} as an AdS to AdS domain wall and to the solution interpolating from \eqref{E:AdSUV} to \eqref{E:LifshitzIR} as an AdS to Lifshitz domain wall. 

For an AdS to AdS domain wall geometry both $G$ and $g$ approach their trivial infrared (small $r$) value via a power law behavior, 
\begin{equation}
\label{E:GglimitAdS}
	G_0 \to p_0 r^{\Delta_G - 3}\,,
	\qquad
	g \to g_0 r^{\Delta_G - 3}\,,
\end{equation}
where $\Delta_G$ is the conformal dimension of the current operator and can be obtained explicitly by considering linearized perturbations of the solution around the infrared background \eqref{E:AdSIR}. Likewise, from \eqref{E:LifshitzIR} and \eqref{E:Lifshitzfluc} we have, at small $r$,
\begin{equation}
\label{E:GglimitLisfhitz}
	G_0 \to p_0 r^z\,,
	\qquad
	g \to g_0 r^z\,,
\end{equation}
for an AdS to Lifshitz domain wall.
Thus, for both an AdS to AdS domain wall and for an AdS to Lifshitz domain wall we find that the ratio $G_0/g$ is finite in the infrared concluding our argument that the charge density of the normal component vanishes at zero temperature.

We note in passing that for a given potential $V(\psi)$ the preferred infrared behavior will be the most stable one. A criteria for stability of the AdS to AdS domain wall is the infrared conformal dimension of the current dual to the gauge field, $\Delta_G$. When this current is relevant in the infrared, $\Delta_G<4$, then the infrared geometry is unstable. The conformal dimension of the current dual to the gauge field can be obtained by considering small fluctuations $\delta G_{IR}$ of the zero component of the gauge field around the background solution \eqref{E:AdSIR}. One finds
\begin{equation}
\label{E:Geq}
	\delta G_{IR}'' + \frac{3}{r} \delta G_{IR}' - \frac{m_{\phi}^2 L_{IR}^2}{r^2} \delta G_{IR} = 0\,,
\end{equation}
where
\begin{equation}
	m_{\phi}^2 = \frac{2 q^2 |\psi_{IR}|^2 V_{\psi}(\psi_{IR})}{V_F(\psi_{IR})} \,.
\end{equation}
Using the standard relation between mass and conformal dimension (see e.g., \cite{D'Hoker:2002aw}), we find that
\begin{equation}
	\Delta_G = 2 + \sqrt{1+ m_{\phi}^2 L_{IR}^2}\,.
\end{equation}	
Thus, AdS to AdS domain walls will certainly be unstable for $\Delta_G \leq 4$ though instabilities may arise even when $\Delta_G>4$ if other competing solutions exist \cite{Gubser:2009cg}.

\section{Numerics}
\label{S:Numerics}

In order to construct the AdS to AdS and AdS to Lifshitz domain wall solution and exhibit the behavior described by \eqref{E:oddRules} explicitly we resort to numerics. We focus our attention on the particular W-shaped scalar potential
\begin{equation}
\label{E:potential}
V(|\psi |)= m^2 |\psi|^2 +\frac{u}{2} |\psi|^4\,,
\end{equation}
with $m^2<0$ and $u>0$ and 
\begin{equation}
\label{E:potentials}
V_\psi = 1\,,\qquad V_F = 1\,.
\end{equation}
Following \cite{Gubser:2009cg} the action characterized by the above potentials can generate both AdS to AdS domain walls and AdS to Lifshitz domain wall solutions depending on the choice of parameters $\{q, m, u\}$. 
When the boundary theory flows to a conformal fixed point (i.e., the geometry is given by an AdS to AdS domain wall), then, following the notation and analysis of section \ref{S:lowT},
\begin{equation}
	\psi_{IR} = \sqrt{-\frac{m^2}{u}}\,,
	\qquad
	L_{IR} = \sqrt{\frac{24 u}{m^4+24 u}}\,.
\end{equation}
When the small $r$ behavior is Lifshitz we find that
\begin{equation}
	\psi_0 = \sqrt{-\frac{m^2}{u} + \frac{2 q^2(z-1)}{z u}}\,,
	\qquad
	\frac{1}{2} u \psi_0^4 + \left( m^2 + \frac{2 q^2(9+z(2+z))}{3 z} \right) - 12 = 0\,,
\end{equation}
where the last equation determines $z$ in terms of $u$, $m$ and $q$. 

While not directly related to our current analysis, the condition that the current dual to the gauge field is irrelevant in the infrared, $\Delta_G>4$, amounts to
\begin{equation}
\label{E:conformal2}-\frac{48\, q^2\, m^2}{24\, u +m^4}> 3\,.
\end{equation}
As observed in \cite{Gubser:2009cg} even when  $\Delta_G > 4$, a solution with Lifshitz geometry in the infrared (small $r$) may have lower free energy than the solution which is asymptotically AdS at small $r$.\footnote{Indeed, for the case at hand if $q<\sqrt{-m^2/3}$ and $1<z<z_+$ or $\sqrt{-m^2/3}<q<\sqrt{-27 m^2/71}$ and $z_-<z<z_+$ where $z_{\pm} = \frac{9m^2+7q^2 \pm \sqrt{81m^4+186 m^2 q^2 - 71 q^4}}{2 m^2 - 4 q^2}$ then both an AdS solution and a Lifshitz solution are possible in the deep interior of the geometry. The preferred solution will be the one with the lowest free energy.}

To compute the integrals in \eqref{E:oddRules} explicitly, we have constructed, numerically, superfluid solutions to the equations of motion at consecutively low temperatures. Our numerical scheme involved integrating \eqref{E:EOMs} subject to the asymptotic boundary conditions given in \eqref{E:UV}, i.e., we imposed that the non normalizable mode of the scalar field vanishes. With numerical expressions for  the background metric and matter fields at hand we solved the equations of motion  \eqref{E:ggamma} for fluctuations of the metric and gauge field, $g$ and $\gamma$, demanding that their near boundary behavior is dictated by \eqref{E:asymg}.  By varying the charge of the scalar field $q$ and its potential which is defined by $m$ and $u$ (see \eqref{E:potential}) we could generate condensates whose low temperature behavior is either an AdS to AdS domain wall or an AdS to Lifshitz geometry. 

\begin{figure}[hbt]
\centering{
\includegraphics[width=0.495\textwidth]{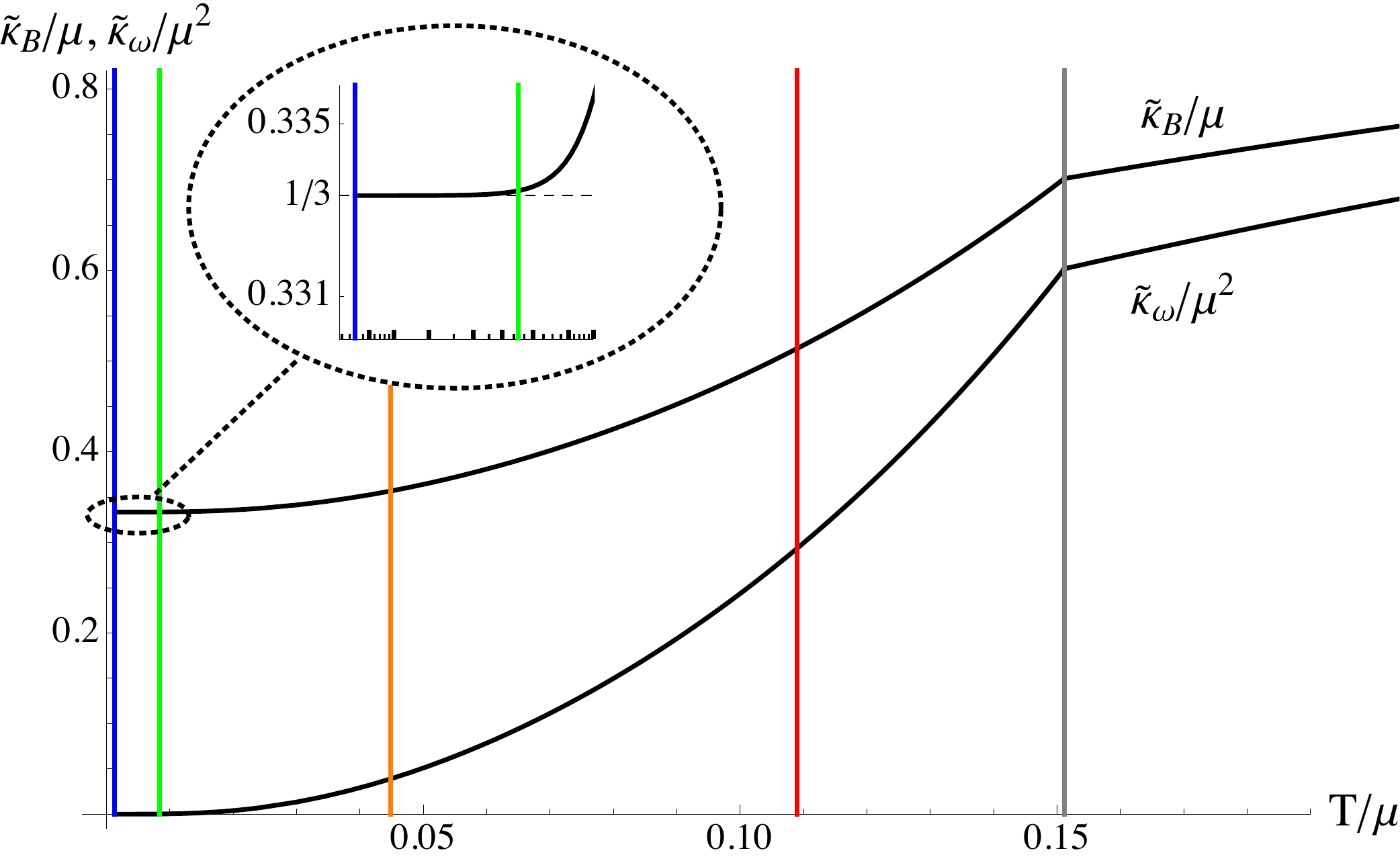}\hfill
\includegraphics[width=0.495\textwidth]{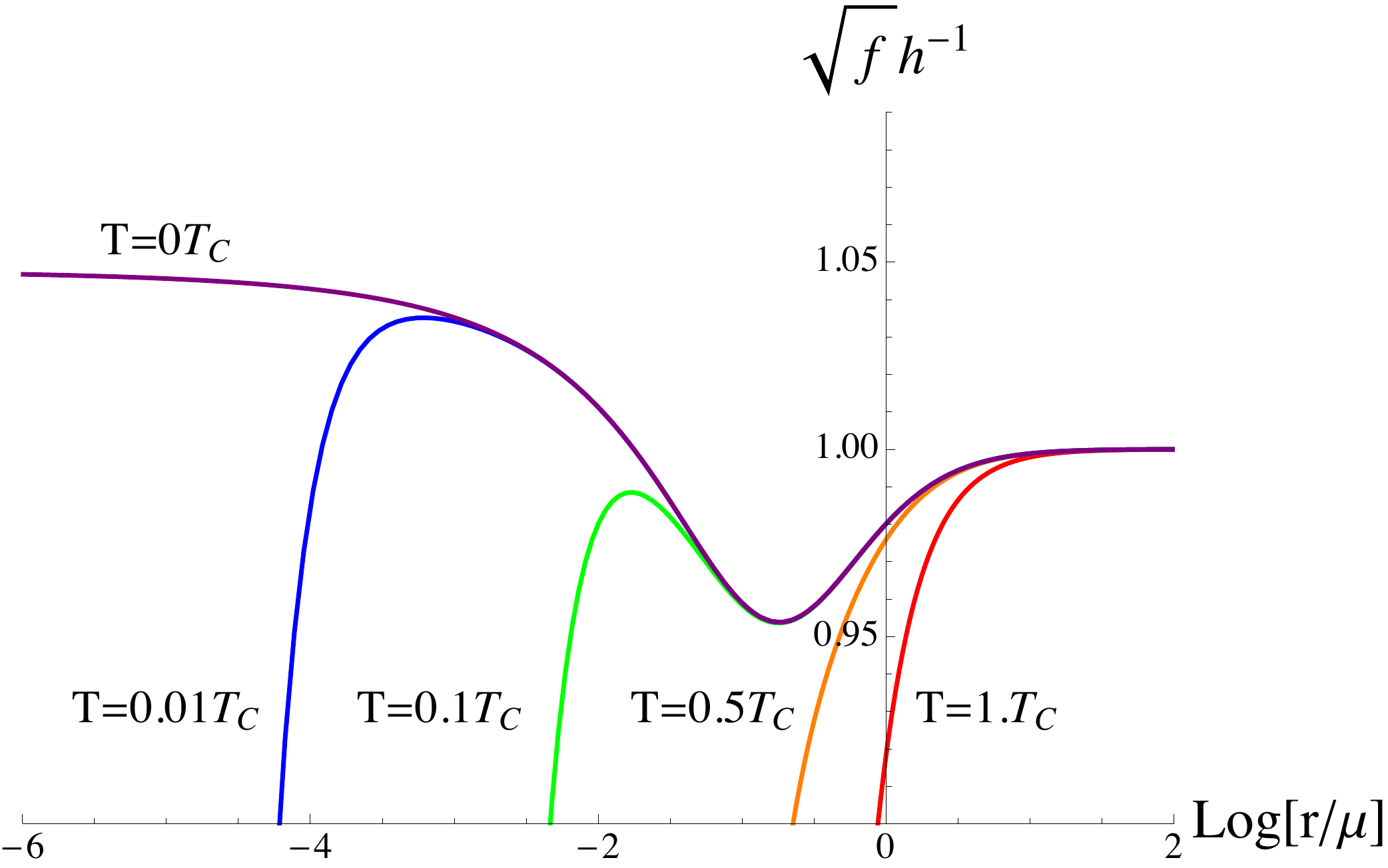}
\caption{\label{F:AdS} A typical AdS to AdS domain wall solution (right) and the associated values of $\tilde{\kappa}_B$ and $\tilde{\kappa}_{\omega}$ (left). In the right panel we have plotted the metric components $\sqrt{f}/h$ (see \eqref{E:ansatz}) for a condensed phase obtained  by solving \eqref{E:EOMs} with $c=1$, $q=2$, \eqref{E:potential} with $m^2 L^2 = -15/4$ ($\Delta=5/2$) and $u=6$, and \eqref{E:potentials}. The zero temperature solution (purple) interpolates between an AdS geometry at the boundary (large $r$) and an AdS geometry in the deep interior (small $r$). In the left panel we plot the chiral  conductivities $\tilde{\kappa}_B$ and $\tilde{\kappa}_\omega$ as a function of temperature. The vertical gray line signifies the critical temperature at which the superfluid phase appears. The remaining red, orange, green and blue vertical lines correspond to the temperatures exhibited in the right panel. We have also added an inset featuring the low temperature behavior of $\tilde{\kappa}_B$, where the horizontal axis (temperature) is given in a logarithmic scale.
}
}
\end{figure}

\begin{figure}[hbt]
\centering{
\includegraphics[width=0.495\textwidth]{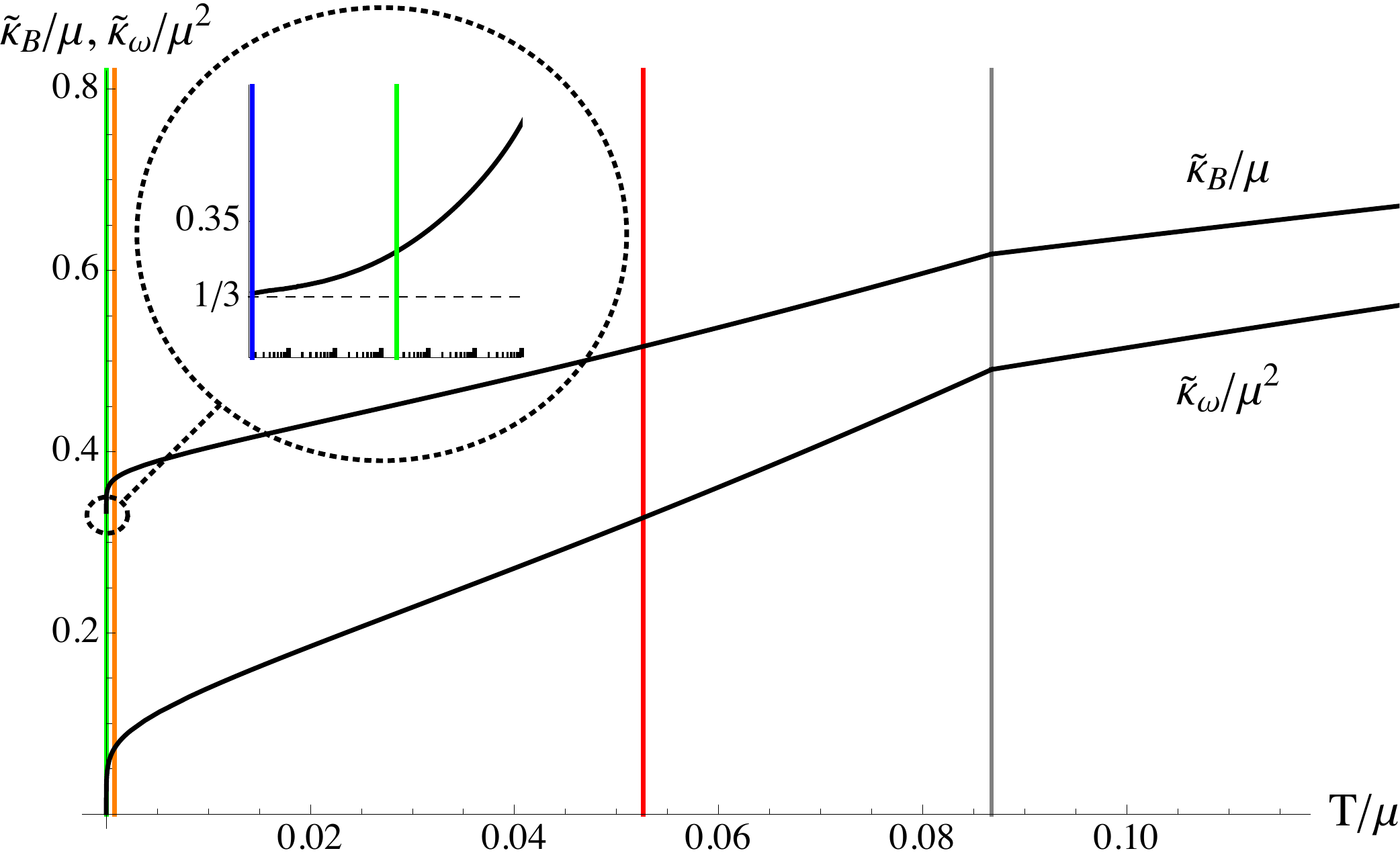}\hfill
\includegraphics[width=0.495\textwidth]{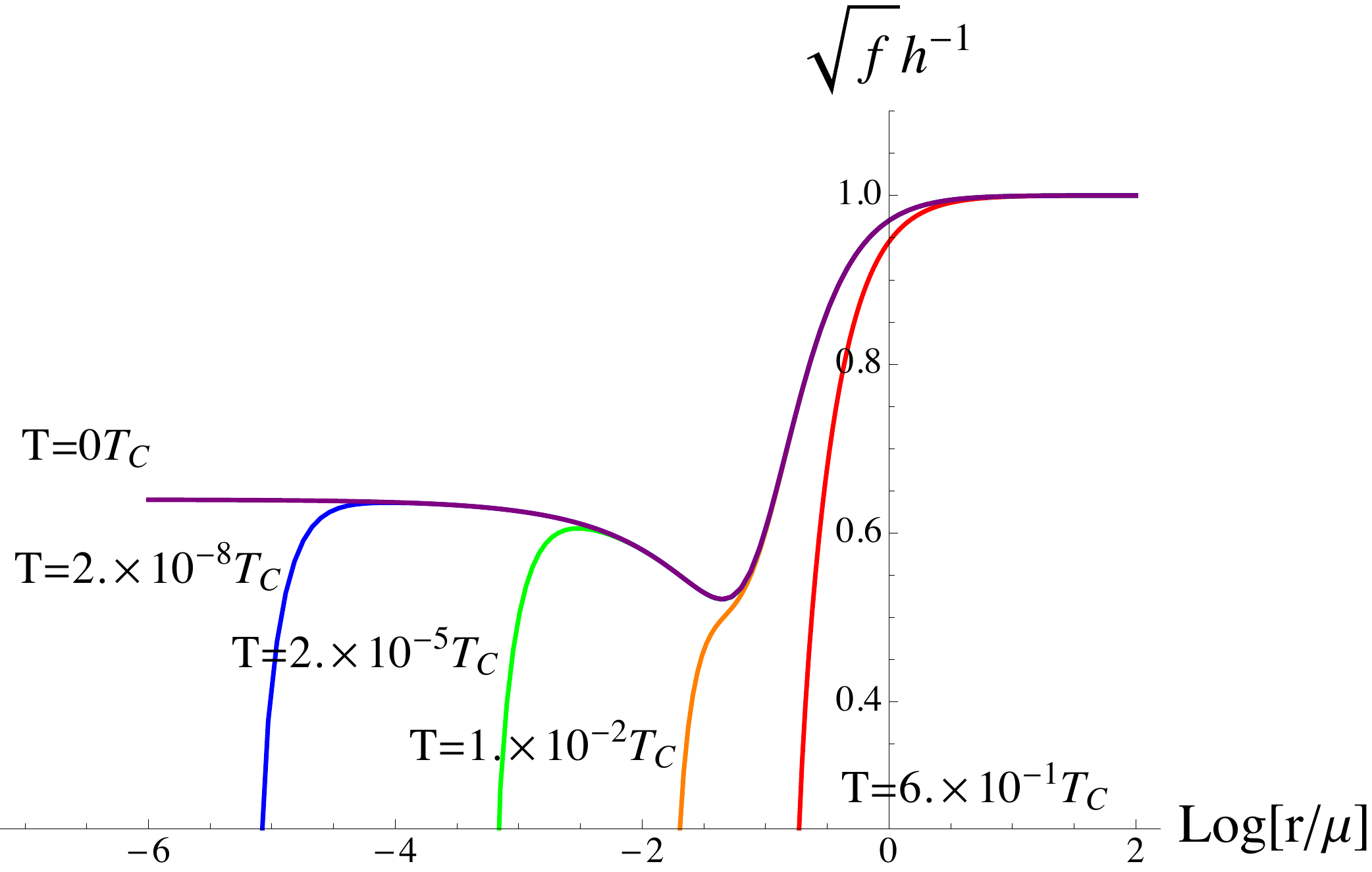}
\caption{\label{F:Lif1} A typical AdS to Lifshitz domain wall solution (right) and the associated values of $\tilde{\kappa}_B$ and $\tilde{\kappa}_{\omega}$ (left). In the right panel we have plotted the metric components $\sqrt{f}/h$ (see \eqref{E:ansatz}) for a condensed phase obtained  by solving \eqref{E:EOMs} with $c=1$, $q=3/2$, \eqref{E:potential} with $m^2 L^2 = -15/4$ ($\Delta=5/2$) and $u=7$, and \eqref{E:potentials} which corresponds to a critical exponent $z=3.68$. The zero temperature solution (purple) interpolates between an AdS geometry at the boundary (large $r$) and a Lifshitz geometry in the deep interior (small $r$). In the left panel we plot the chiral  conductivities $\tilde{\kappa}_B$ and $\tilde{\kappa}_\omega$ as a function of temperature. The vertical gray line signifies the critical temperature at which the superfluid phase appears. The remaining red, orange, green and blue vertical lines correspond to the temperatures exhibited in the right panel. We have also added an inset featuring the low temperature behavior of $\tilde{\kappa}_B$, where the horizontal axis (temperature) is given in a logarithmic scale.}
}
\end{figure}

Plots exhibiting the emergence of domain wall geometries can be found in the right panel of figures \ref{F:AdS} and \ref{F:Lif1} where the metric coefficients $f$ and $h$ defined in \eqref{E:ansatz} are plotted as a function of the radial coordinate $r$ for various values of the temperature. The purple curve in the right panel of figure \ref{F:AdS} depicts a typical AdS to AdS domain wall geometry. 
The blue, green, orange and red curves depict the AdS to AdS domain wall solution at increasingly higher temperatures.  The domain wall geometry becomes imperceptible once the temperature is larger than, roughly $10^{-3}T_c$ with $T_c \sim 0.15 \mu$. In contrast, for an AdS to Lifshitz geometry to be revealed one has to reach significantly lower temperatures. In the right panel of figure \ref{F:Lif1} an emergent Lifshitz symmetry at small values of $r$ is observed at temperatures smaller than, roughly, $10^{-9}T_c$.
In general, we find that the temperature at which a Lifshitz geometry in the infrared becomes manifest is lower the higher the critical exponent.

In our analysis we have computed the values of the chiral conductivities for a scalar of mass $m^2 L^2 = -15/4$ which corresponds to an operator of dimension $\Delta = 5/2$ or, using the alternative quantization scheme \cite{Klebanov:1999tb}, $\Delta = 3/2$.
For the $\Delta=5/2$ case we have generated 17 solutions with values of $q$ and $u$ in the range $3/2<q<6$ and $5.1<u<33$. We have also studied a handful of solution for the alternate quantization scheme where $\Delta=3/2$.
Typical behavior of the chiral conductivities $\tilde{\kappa}_\omega$ and $\tilde{\kappa}_B$ for configurations which reduce to an AdS to AdS geometry at low temperatures can be seen in the left panel of figure \ref{F:AdS}. We have not shown very similar plots for the coefficients $\sigma_\omega$ and $\sigma_B$ associated with the entropy current. The chiral conductivities converge monotonically to their universal low temperature values \eqref{E:asymptotics} reaching $1\%$ accuracy at roughly $T \sim 0.3 T_c$ and $0.01\%$ accuracy at $T \sim 0.1 T_c$. We attribute these considerably rapid convergence properties to the appearance of the domain wall solution at relatively high temperatures. 

Typical behavior of the chiral conductivities for a configuration which reduces to an AdS to Lifshitz domain wall can be found in the left panel of figure \ref{F:Lif1}. Here too the chiral conductivities approach their universal values monotonically with decreasing temperature but in contrast to the AdS to AdS domain wall, convergence becomes good only for very cold configurations; for the solution in figure \ref{F:Lif1}, $\tilde{\kappa}_B$ and $\tilde{\kappa}_{\omega}$ approach their universal value as given in \eqref{E:asymptotics} with $1\%$ accuracy at temperatures $T\sim 10^{-6}T_c$. For  smaller values of the critical exponent, $z = 2$, we found that the chiral conductivities approached their universal value to within $1\%$ at $T \sim 10^{-4} T_c$. It would be interesting to obtain a precise relation between the rate of convergence of the chiral conductivities to their universal value and the value of the critical  exponent $z$ of the underlying infrared Lifshitz theory.

Our main result \eqref{E:asymptotics} seems to be robust and raises the question on its validity beyond the holographic regime. One possibility is that \eqref{E:asymptotics} are a generic result valid in holographic theories but valid more generally, much like the shear viscosity to entropy ratio \cite{Policastro:2001yc} or a linear combination of second order transport coefficients \cite{Haack:2008xx,Shaverin:2012kv}. Another possibility is that the relations \eqref{E:asymptotics} are bona fide and apply to superfluids on a more general level similar to the relations in \eqref{E:kappatildenormal} which were also discovered holographically \cite{Erdmenger:2008rm} and then understood more generally \cite{Son:2009tf}. Possible future directions which may test which of these two possibilities is the correct one include an analysis similar to the one carried out here but in a different number of space-time dimensions, a computation of the effects of  other types of anomalies such as the mixed anomaly, or studying the effect of other types of parity breaking terms in the bulk action which do not generate an anomaly. We leave such investigations for the future.

\section*{Acknowledgments}
We thank S. Bhattacharyya, C. Hoyos and K. Jensen for comments on a previous version of this manuscript. IA is partly supported by the Lady Davies Foundation. AY is a Landau fellow, supported in part by the Taub foundation. IA, NL and AY are also supported by the ISF under grant number 495/11, by the BSF under grant number 2014350, by the European commission FP7, under IRG 908049 and by the GIF under grant number 1156-124.7/2011.

\begin{appendix}

\end{appendix}	

\bibliographystyle{JHEP}
\bibliography{ZeroTChiralbib}

\end{document}